\documentclass[traditabstract]{aa}
\usepackage{txfonts}
\usepackage{graphicx}
\usepackage{natbib}
\bibpunct{(}{)}{;}{a}{}{,} 

\usepackage{xspace}
\newcommand{\cyd}{\ensuremath{\,\mathrm{d}^{-1} }\xspace} 

\begin{document}

\title{The Discovery of Stellar Oscillations in the K Giant $\iota$~Dra}

\author{Mathias Zechmeister\inst{1,2} \and
        Sabine Reffert\inst{3} \and
        Artie P. Hatzes\inst{2} \and
        Michael Endl\inst{4} \and
        Andreas Quirrenbach\inst{3}}

\institute{Max-Planck-Institut f\"{u}r Astronomie, K\"{o}nigstuhl 17, 69117 Heidelberg, Germany
  \and Th\"{u}ringer Landessternwarte Tautenburg (TLS), Sternwarte 5, 07778 Tautenburg, Germany
  \and ZAH-Landessternwarte, K\"{o}nigstuhl 12, 69117 Heidelberg, Germany
  \and McDonald Observatory, University of Texas, Austin, TX 78712, USA
}

\date{Received; accepted}

\abstract{$\iota$~Dra (\object{HIP~75458}) is a well-known example for a K giant hosting a substellar companion since its discovery by \citeauthor{Frink02} \citeyearpar{Frink02}. We present radial velocity measurements of this star from observations taken with three different instruments spanning nearly 8 years. They show more clearly that the RV period is long-lived and coherent thus
supporting the companion hypothesis. The longer time baseline now allows for a more accurate determination of the orbit with a revised period of $P=511\,$d and an additional small linear trend, indicative of another companion in a wide orbit. Moreover we show that the star exhibits low amplitude, solar like oscillations with frequencies around 3-4\cyd  (34.7-46.3\,$\mu$Hz).}

\authorrunning{M. Zechmeister et al.}
\maketitle

\section{Introduction}

Up to now long period radial velocity (RV) variations have been detected in several K giants, e.g. $\beta$~Gem \citep{Hatzes06,Reffert06}, HD~47536 \citep{Setiawan03}, HD~13189 \citep{Hatzes05} and $\iota$~Dra \citep{Frink02}.
Rotational modulation can be excluded as a cause of these variations for most of these giants due to lack of variations in photometry or line bisectors with the RV period \citep{Hatzes98Bisector}. The most likely interpretation is orbiting, substellar companions.

Since the progenitor stars to planet hosting giant stars can have masses significantly larger than $1M_{\odot}$, these giant stars offer a way to investigate planet formation around massive stars. In their evolved phase the Doppler induced effects of a planet are easier to detect than in their main-sequence phase, due to cooler effective temperatures and smaller rotational velocities.

But for evolved stars the determination of the mass is more difficult. The most practical way for determining the stellar mass is the use of evolutionary tracks, but those tracks converge in the region of the giant branch for stars over a wide range of masses and thus makes the derived masses uncertain.

Another possibility to measure the stellar mass is via asteroseismology. This requires the investigation of stellar oscillations. Photometric and RV variations with short periods from hours to days have already been detected in some K and late G giants (e.g. $\alpha$~Boo: \citealp{Retter03,Tarrant07}; $\alpha$~Ari: \citealp{Kim06}; $\epsilon$~Oph: \citealp{deRidder06}; $\xi$~Hya: \citealp{Frandsen02}, \citealp{Stello06}; $\beta$~UMi: \citealp{Tarrant08}) and are consistent with solar-like p-mode oscillations. Such oscillations have also been measured for ensembles of red giants with photometry of star rich regions or clusters (e.g. \citealp{Gilliland08}, \citealp{Stello08}) which is a very efficient way to perform such asteroseismologic studies.

Solar like oscillations have also been discovered in planet hosting main sequence stars (e.g. $\mu$~Arae: \citealp{Bazot05}; $\iota$~Hor: \citealp{Vauclair08}) and in the planet hosting K giant $\beta$~Gem for which \cite{Hatzes07} gave, in combination with interferometric measurements of the angular diameter, an estimation for the stellar mass completely independent from evolutionary tracks. In a similar way we will do this here for $\iota$~Dra.

\section{Stellar Parameters}

In Table \ref{tab:Param1} we summarize some direct measurements for the K2III star $\iota$~Dra. The improved parallax from the Hipparcos catalog by \citet{vanLeeuwen07} is given. Angular diameter estimates based on spectrophotometry are available from the CHARM2 catalog \citep{Richichi05}.

\begin{table}[htb!]
\caption{\label{tab:Param1}Stellar parameters of $\iota$~Dra.}
\centering
\begin{tabular}{lc}
\hline\hline 
Parameter& Value\\
\hline 
Spectral type$^{\rm{a}}$                       & K2III\\
$V^{\rm{a}}$ {[}mag{]}                       & $3.29$\\
$M_{V}$ {[}mag{]}                   & $0.81$\\
$L$ {[}$L_{\odot}${]}               & $64.2\pm2.1$\\
Parallax$^{\rm{b}}$ $p$ {[}mas{]}              & $32.23\pm0.1$\\
Distance $d$ {[}pc{]}               & $31.03\pm0.1$\\
Angular Diameter$^{\rm{c}}$  $\theta$ {[}mas{]} & $3.73\pm0.04$\\
Radius $R$ {[}$R_{\odot}${]}        & $12.38\pm0.17$\\
\hline
\end{tabular}
\begin{list}{}{}
\item[$^{\rm{a}}$] \citep{ESA97}
\item[$^{\rm{b}}$] \citep{vanLeeuwen07}
\item[$^{\rm{c}}$] \citep{Richichi05}
\end{list}
\end{table}

\begin{table*}[htb!]
\caption{\label{tab:Param2}Stellar parameters of $\iota$~Dra from literature ($T$, $\log g$ and [Fe/H]) and derived with evolutionary tracks ($M$, $\log g$ and $R$).}
\centering
\begin{tabular}{llllccc}
\hline\hline 
 Reference         & $T$ [K]   &$\log g$ & [Fe/H]                 &$M$ {[}$M_{\odot}${]} & $\log g$ & $R$ {[}$R_{\odot}${]}\\
\hline 
\cite{Soubiran08}  & 4552         & 2.96           & 0.16              & $1.40\pm0.24$   & $2.40\pm0.10$ & $11.82\pm0.63$\\
\cite{Prugniel07}  & 4543         & 2.88           & 0.19              & $1.41\pm0.23$   & $2.40\pm0.10$ & $11.94\pm0.62$\\
\cite{Hekker07}    & 4605         & 2.96           & 0.07              & $1.39\pm0.24$   & $2.45\pm0.10$ & $11.19\pm0.58$\\
\cite{Santos04}    & $4775\pm113$ & $3.09\pm0.40$  & $0.13\pm0.14$     & $1.71\pm0.38$   & $2.61\pm0.12$ & $10.34\pm0.40$\\
\cite{Gray03}      & 4526         & 2.64           & 0.11              & $1.31\pm0.24$   & $2.37\pm0.11$ & $11.97\pm0.65$\\
\cite{Prugniel01}  & 4491         & 2.57           & 0.06              & $1.24\pm0.24$   & $2.32\pm0.10$ & $12.23\pm0.65$\\
\cite{Cenarro01,Cenarro07} & 4498 & 2.38           & 0.05              & $1.24\pm0.24$   & $2.33\pm0.10$ & $12.15\pm0.66$\\
\cite{Prieto99}    & $4466\pm100$ & $2.24\pm0.35$  &                   & $1.05\pm0.36$\\
\cite{McWilliam90} & 4490         & 2.74           & $0.03\pm0.11$     & $1.23\pm0.24$   & $2.32\pm0.10$ & $12.22\pm0.66$\\
\cite{Williams74}  & $4530\pm100$ & $2.60\pm0.25$  & $0.29\pm0.20$     & $1.40\pm0.23$   & $2.39\pm0.10$ & $12.06\pm0.64$\\
\hline
\end{tabular}
\end{table*}

The quantity which cannot be measured for single stars directly is the mass, which always relies on evolutionary tracks. With the online tool\footnote{http://stev.oapd.inaf.it/cgi-bin/param} from Girardi (see \citealp{daSilva06} and \citealp{Girardi00} for a description), we derived some values for $M$, $\log g$ and $R$. This tool uses as input parameters the visual magnitude $V$, parallax $p$, temperature $T$ and metallicity Fe/H. For the latter both quantities, $T$ and [Fe/H], we adopt the values  in
Table~\ref{tab:Param2} as given by several authors. \citet{Prieto99} gave no metallicity but derived themselves a mass of $M=1.05M_\odot$ which was used for the calculation of the companion mass in \citet{Frink02}. This value is at the
lower mass range resulting from Table~\ref{tab:Param2} (1.05 -- $1.71M_\odot$).  This shows how much the tracks depend on temperature and metallicity. Furthermore, Table~\ref{tab:Param2} allows a cross check of the $\log g$-values derived from spectroscopy and from evolutionary tracks. The $\log g$-values derived from spectroscopy are always higher, which illustrates the difficulty of reliable mass determinations for K giants.

\section{Observations}

For our analysis we have RV measurements for $\iota$~Dra from three independent instruments (Table~\ref{tab:Journal-of-observations} and Figure~\ref{fig:RV}). The data set from the 0.6m CAT (Coud\'{e} Auxiliary Telescope) with the Hamilton \'{E}chelle Spectrograph at Lick Observatory provides nearly 8 years and thus the longest time baseline for the orbit determination.
The data up until March 2002 were already published in \citet{Frink02}.

At the Thuringia State Observatory (TLS -- Th\"{u}ringer Landessternwarte) $\iota$~Dra was observed on 43 nights spanning 200$\,$d. The Alfred-Jensch 2m telescope is equipped with a Coud\'{e} \'{E}chelle Spectrograph with a wavelength coverage from 4630 to 7370\,\AA.
These observations primarily were carried out to look for stellar oscillations with short periods. For this program on some nights up to 20 spectra (even 61 on the night of JD=2454193) were taken (Figure~\ref{fig:RVtls}).

With a rather short time baseline of 8 days some spectra were obtained in June 2005 with the 2.7$\,$m telescope at the McDonald Observatory and the ``2dcoude'' spectrograph. This instrument provides a wavelength range from 3600\,{\AA} to 1$\,\mu$m. The RVs were calculated using the ``Austral'' program \citep{Endl00}. The whole data set is plotted in Figure \ref{fig:RVmcd}.

All three instruments have a resolving power around $R=\frac{\lambda}{\Delta\lambda}=60\,000$ and utilize an iodine absorption cell for the wavelength reference. Table~\ref{tab:Journal-of-observations} lists the time coverage, time span $T$, number of spectra $N$ and average precision of the radial velocity measurements $\sigma_{RV}$. These measurement errors are internal errors due to instrumental effects recorded with the iodine lines. Comparing with point to point scatter, e.g. in Figure~\ref{fig:RVtls}, the errors appear reasonable.

\begin{table}[htb!]
\caption{\label{tab:Journal-of-observations}Journal of observations.}
\centering
\begin{tabular}{ccrrc}
\hline\hline
Data Set&Coverage&\multicolumn{1}{c}{$T$ {[}d{]}}&
\multicolumn{1}{c}{$N$}&$\sigma_{RV}$ {[}m/s{]}\\
\hline
CAT & 2000.39--2008.38 & 2918 & 147 & 4.3\\
TLS & 2006.99--2007.59 & 221  & 280 & 3.3\\
McD & 2005.44--2005.46 & 8    & 62  & 3.9\\
\hline
\end{tabular}
\end{table}

\section{Orbital Solution}

The parameters of the orbit were determinated by weighted least squares ($\chi^2$-Fit) with the differential correction method \citep{Sterne41}.
We fitted simultaneously five Keplerian orbital elements, a linear trend and three offsets for the combined data sets as the
measurements give relative velocities and the three instruments have different zero points. When weighting the measurements one has to take into account that not only the measurement errors introduce an error into the parameter determination but also the jitter
due to pulsations: $\sigma_i^2=\sigma_{RV,i}^2+\sigma_{P}^2$. The stellar oscillations are discussed in the next chapter but we mention here that $\sigma_P$ is of the order of $10\,$m/s. We adopt this value which is larger than the typical measurement error and therefore leads to a more equal weighting of all measurements in our fit.

\begin{figure*}[htb!]
\centering\includegraphics[width=1\linewidth]{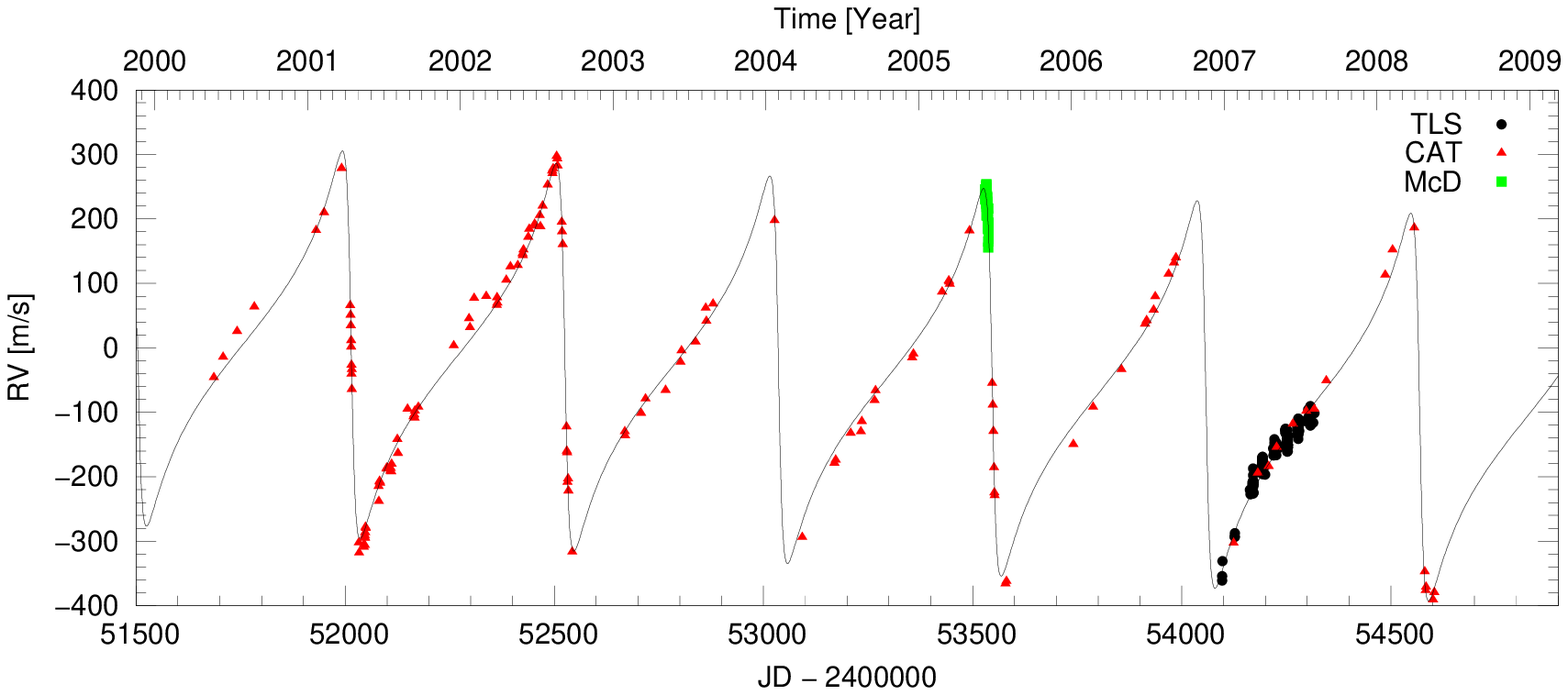}
\centering\includegraphics[width=1\linewidth]{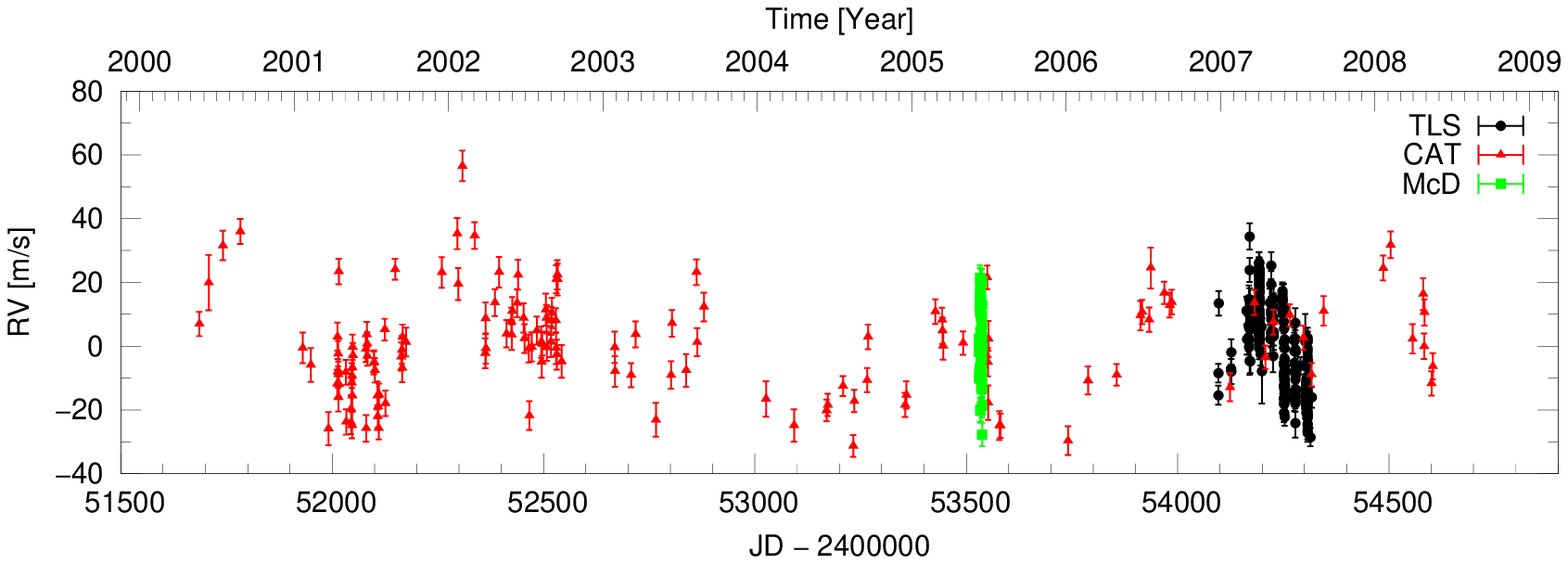}

\caption{\label{fig:RV}Radial velocity measurements for $\iota$~Dra from three data sets and the orbital solution as solid line.
The lower panel shows the residuals after subtraction of the orbital solution and the linear trend. (See online version for a colored figure.)}
\end{figure*}

The resulting orbital elements are listed in Table \ref{tab:orbit} and the RV curve is plotted in Figure~\ref{fig:RV}. In the old
solution the short time baseline and the unrecognizable linear trend had resulted in an overestimated period ($P=536\pm 6\,$d:
\citealp{Frink02}). The linear trend is $-13.8\pm 1.1\,$m/s/yr and may be part of a longer period caused by a further companion. For the calculation of the companion mass we assumed a stellar mass of $1.4M_{\odot}$ (33\% higher than in \citealp{Frink02}).

\begin{table}[htb!]
\caption{\label{tab:orbit}Orbital parameters for the companion to $\iota$~Dra.}
\centering
\begin{tabular}{lc}
\hline\hline 
Parameter&Value\\
\hline
Period $P$ {[}d{]}                                & 510.88 $\pm$ 0.15\\
Amplitude $K$ {[}m/s{]}                           & 299.9  $\pm$ 4.3\\
Periastron time $T_{0}$ {[}JD{]}                  & 2452013.94 $\pm$ 0.48\\
Longitude of periastron $\omega$ {[}\degr{]}      & 88.7 $\pm$ 1.4\\
Eccentricity $e$                                  & 0.7261 $\pm$ 0.0061\\
Linear trend {[}m/s/yr{]}                         & -13.8  $\pm$ 1.1 \\
Mass function$^{\rm a}$ $f(m)$ {[}$M_{\odot}${]}  & $(4.64\pm 0.33) \cdot 10^{-7}$\\
Semi-major axis $a$ {[}AU{]}                      & 1.34\\
companion mass $m\sin i$ {[}$M_{\mathrm{Jup}}${]} & 10.3\\
\hline
\end{tabular}
\begin{list}{}{}
\item[$^{\rm{a}}$] $f(m)=\frac{(m\sin i)^3}{(M+m)^2}=\frac{P}{2\pi G}(K\sqrt{1-e^2})^3$
\end{list}
\end{table}

\begin{figure}[htb!]
\centering\includegraphics[width=1\linewidth]{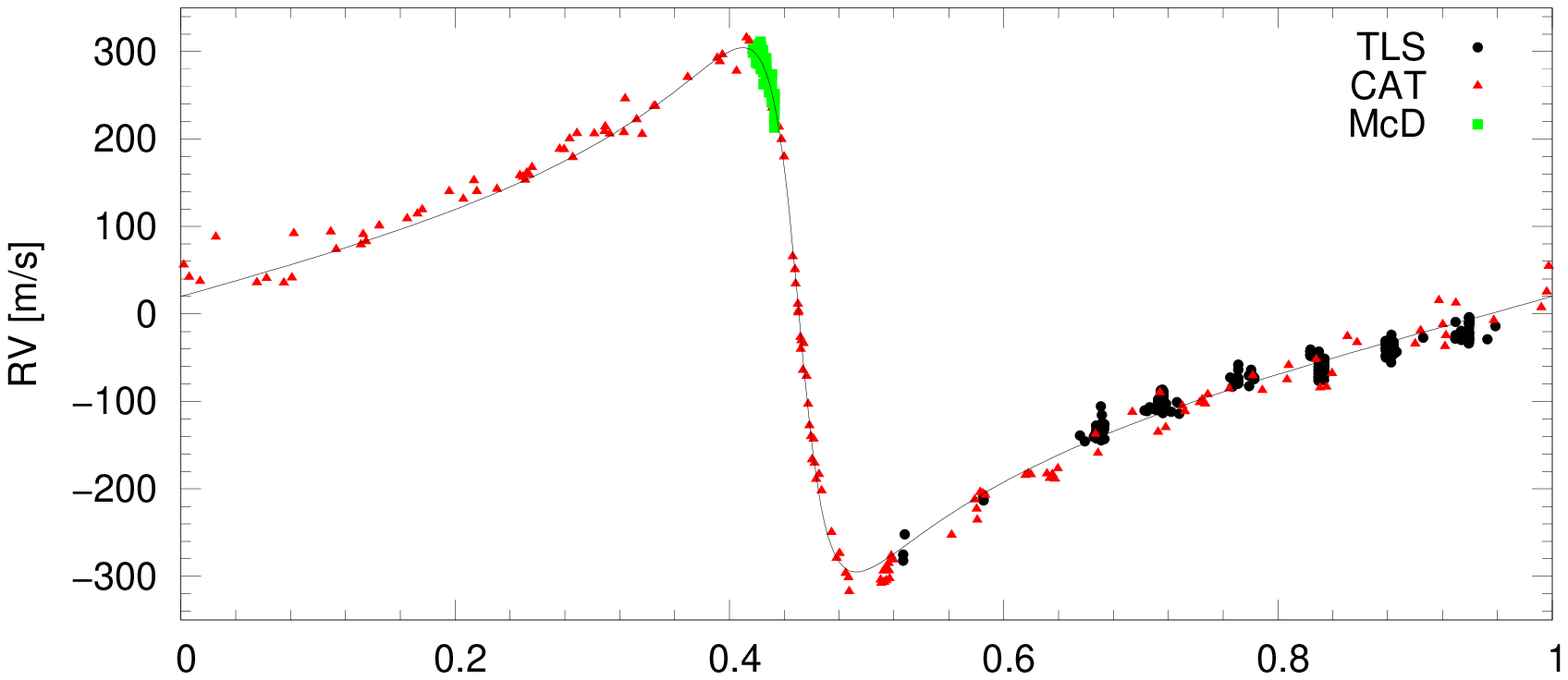}
\centering\includegraphics[width=1\linewidth]{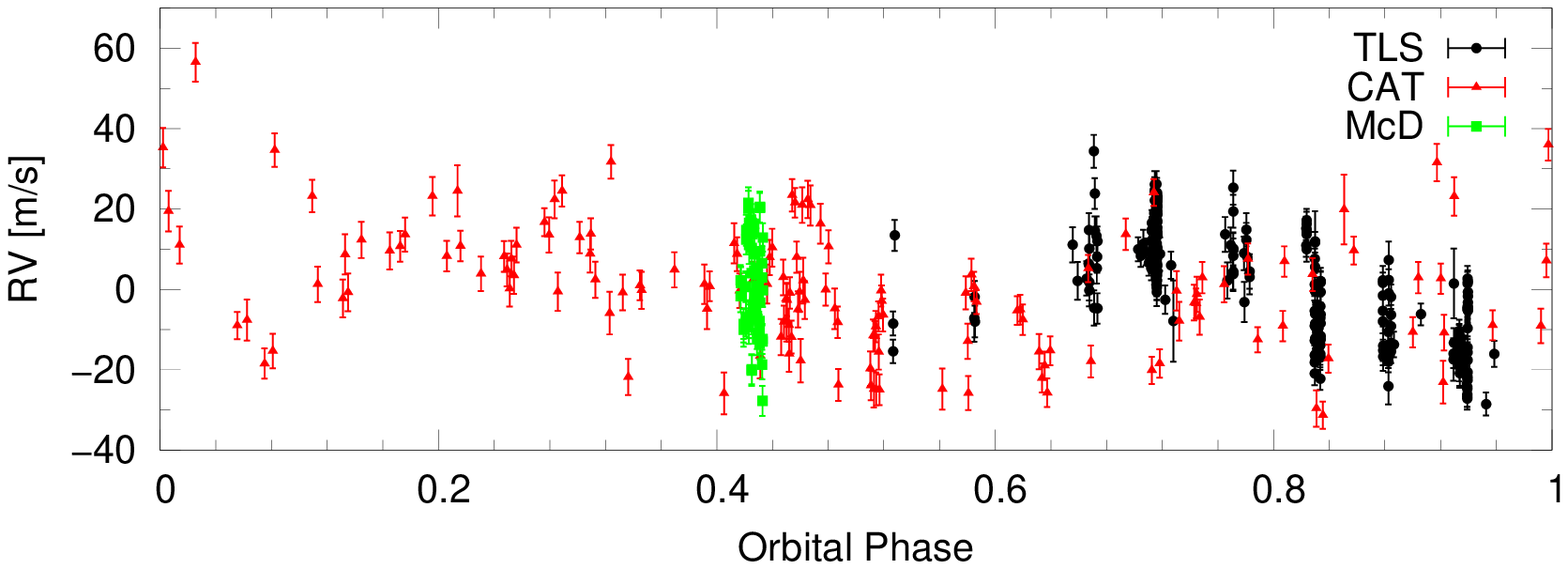}

\caption{\label{fig:phasedRV}Radial velocity measurements for $\iota$~Dra from three data sets phased to the orbital period. The lower panel shows the orbit residuals.  (See online version for a colored figure.)}
\end{figure}

Figure \ref{fig:phasedRV} illustrates the radial velocity phased to the orbital period. The total remaining scatter is 13.9$\,$m/s for all data sets (and for the individual sets, which cover very different time scales: 15.4$\,$m/s (CAT), 13.9$\,$m/s (TLS) and 11.4$\,$m/s (McD)).
This scatter is a factor of 3-4 greater than the measurement errors given in Table \ref{tab:Journal-of-observations} and is largely due to stellar oscillations as we will explain below.

\section{Short Period Oscillations}

\begin{figure}[htb!]
\centering\includegraphics[width=1\linewidth]{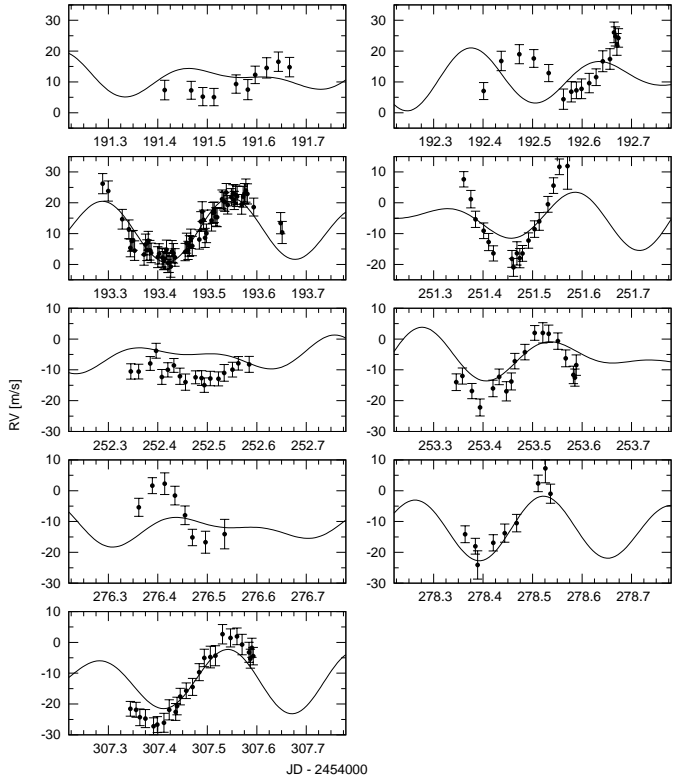}

\caption{\label{fig:RVtls}Radial velocity during several good nights measured at TLS (orbit subtracted). The solid curve is the fit with the TLS frequencies ($f_0$, $f_1$ and $f_2$).}
\end{figure}

\begin{figure}[htb!]
\centering\includegraphics[width=1\linewidth]{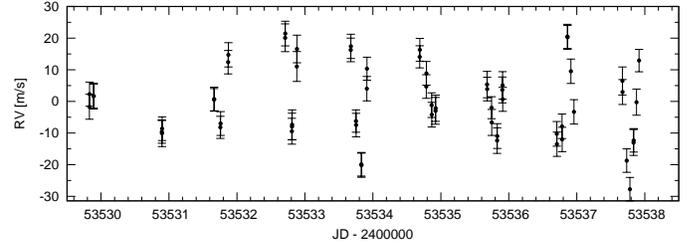}

\caption{\label{fig:RVmcd}Radial velocity during 8 consecutive nights measured at McD (orbit subtracted).}
\end{figure}

For the analysis of variations on short time scales we subtracted the trend and orbit from the data sets. The remaining residuals during some nights for the TLS and McD data sets are shown in Figures \ref{fig:RVtls} and \ref{fig:RVmcd}, respectively. Clearly one can see RV variations with an amplitude of few 10\,m/s.

We performed the frequency analysis of the orbit residuals for each data set individually due to their different time sampling.
For this we used the prewhitening procedure provided by the program Period04 \citep{Lenz04}. A Fourier transformation was performed for the data, the dominant frequency extracted and subtracted from the data. This procedure can be repeated on the subsequent residuals until no further significant frequencies can be found. The prewhitening has the advantage that confusing alias peaks (here mainly 1 day aliases) of the extracted frequency are removed from periodograms.

\begin{figure}[htb!]
\centering\includegraphics[width=1\linewidth]{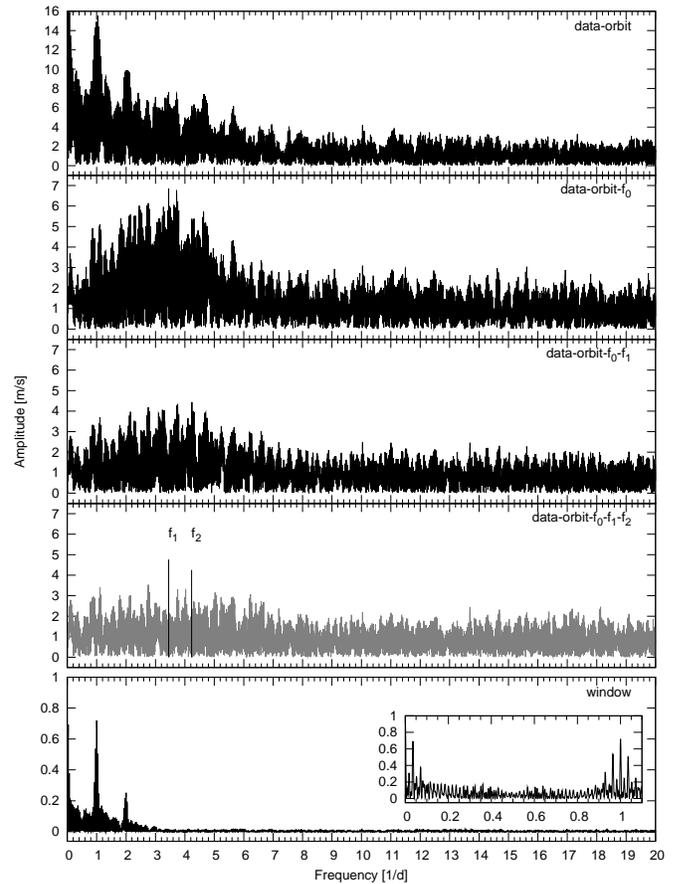}

\caption{\label{fig:tls_f}Amplitude periodograms of the TLS RV data at each prewhitening step. The fourth panel shows the periodogram of the residuals in comparison with the extracted frequencies and the last panel shows the window function.  The inset is an enlargement where the one month alias can be seen.}
\end{figure}

\begin{figure}[htb!]
\centering\includegraphics[width=1\linewidth]{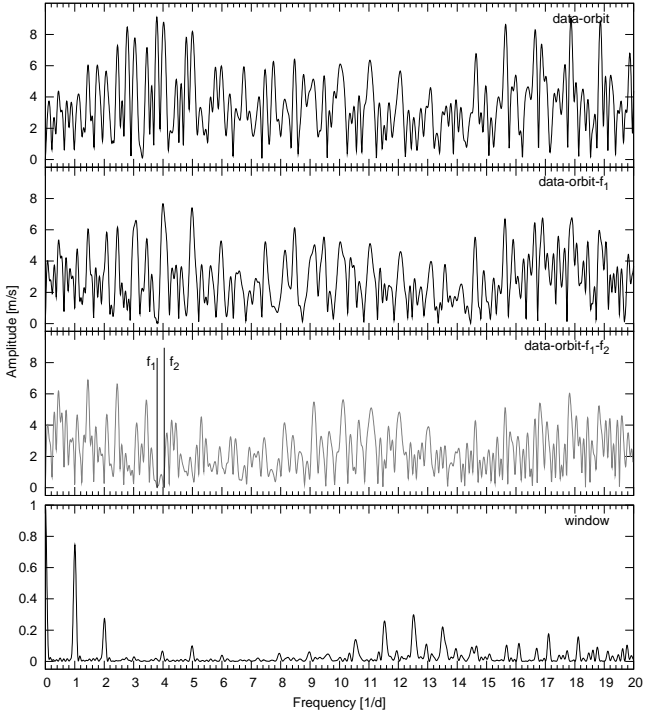}

\caption{\label{fig:mcd_f}Amplitude periodograms of the McD RV data at each prewhitening step. The third panel shows also the extracted frequencies and the last panel shows the window function.}
\end{figure}

Figures~\ref{fig:tls_f} and \ref{fig:mcd_f} show the periodograms in each prewhitening step for the TLS and McD residuals, respectively. The TLS residuals show peaks at 0.040\cyd and 0.004\cyd , which seem to be 1\,month-aliases ($\Delta f=0.036\cyd$) to each other. Neither frequency is found in the other data sets. However, the corresponding long periods ($P=25$\,d and $P=250$\,d) are not of further interest for our investigation of short period oscillations. Subtraction of the lower frequency ($f_0=0.004\cyd$, 12.33\,m/s) flattens the TLS orbit residuals and removes the 1\,day aliases  of $f_0$ (second panel in Figure \ref{fig:tls_f}). This prewhitening step reveals effectively an excess at frequencies around 3.8\cyd. Extraction of the two frequencies $f_{1}$=3.45\cyd (4.77\,m/s) and $f_{2}$=4.23\cyd (4.26\,m/s) lowers the excess power considerably. A third frequency $f_3=$3.75\cyd (3.86\,m/s) is already near an amplitude signal to noise ratio of  S/N=4 (the mean noise level is 0.865\,m/s in the range of 10-20\cyd). This threshold, as suggested by \citet{Kuschnig97}, can be used as a criterion for stopping the prewhitening procedure.

The fit with the frequencies is also drawn in Figure \ref{fig:RVtls}. Nights with many data points fit well, while some discrepancies exist for nights with sparse measurements. This may indicate that more frequencies are present or that the modes
have a finite lifetime due to stochastic excitation mechanisms \citep[e.g][]{Barban07}.

For the McD orbit residuals an excess of  power (upper panel in Figure \ref{fig:mcd_f}) can be seen in the same region as for the TLS data. The peaks at frequencies $>$15\cyd in the McD data set are due to the sparser sampling leading to a stronger aliasing at higher frequencies. 
Extracted are two frequencies, $f_{1}=$3.81\cyd (8.45\,m/s) and  $f_{2}$=4.03\cyd (8.90\,m/s), which are also illustrated along with the periodogram of the remaining residuals in the third panel of figure \ref{fig:mcd_f}.

The higher amplitudes of the frequencies in the McD data may be an effect of the short time baseline and the finite mode lifetime of solar-like oscillations seen in main sequence stars and red giants \citep{Stello06,Carrier07}.

The CAT residuals show a 555\,d period, which is close to the orbital period. However the data set has insufficient time resolution to look for short period variability ($<1$ day) since its Nyquist frequency is only 0.5\cyd.

\section{Discussion}

We compare our results with the scaling relations for solar-like oscillations from \citet{Kje95}. The oscillation velocity amplitude $v_{\mathrm{osc}}$ and the frequency $f_{\mathrm{max}}$ of the strongest mode of the 5-minute-oscillations in the sun scales to other stars as:
\begin{equation}
v_{\mathrm{osc}}=\frac{L/L_{\odot}}{M/M_{\odot}}\cdot0.234\,\mathrm{m/s}
\end{equation}
\begin{equation}
f_{\mathrm{max}}=62\cyd\frac{(T/5777K)^{3.5}}{v_{\mathrm{osc}}/\,\mathrm{m\, s}^{-1}}.\label{eq:fmax}
\end{equation}
These relations are valid for a wide range of convective stars \citep{Bedding03} and seem to be valid for other K giants \citep{Stello08}. For a power excess frequency of around 3.8\cyd, which is present in two data sets, the expected amplitude should be 6.75$\,$m/s according eq.(\ref{eq:fmax}) (based on $T=4490\,\mathrm{K}$, \cite{McWilliam90}). This agrees with the excess power seen in the periodograms and is comparable to the amplitude of $f_1$ (4.77$\,$m/s (TLS) and 8.90$\,$m/s (McD)). So this frequency satisfies the relation between amplitude and frequency typical for solar-like oscillations.

Under the assumption that the simple scaling relations are valid, the combination of both equations to $L/M\propto T^{3.5}/f_\mathrm{max} $ implies a luminosity to mass ratio of $L/M=29L_{\odot}/M_{\odot}$  for this frequency. Using $L=64.2\pm2.1L_{\odot}$ from the Hipparcos catalog one would estimate a mass of $2.2M_{\odot}$ (or with $f_{\mathrm{max}}\propto g/\sqrt{T}$ a $\log g$ of 2.54).
Even if we roughly estimate an uncertainty of $\pm0.6M_\odot$ due to strong 1 day aliasing (i.e.\ derived from $\Delta f_{\mathrm{max}}=\pm1$\cyd), this value seems too high in comparison with Table \ref{tab:Param2}. For a mass of $1.4M_{\odot}$ the expected frequency is 2.4\cyd. We cannot exclude that we miss such a mode as 0.42$\,$d periods are difficult to detect in short nights with single site observations.

A more accurate mass estimation could be done with the scaling relation for the frequency splitting:
\begin{equation}
\Delta f_{0}=\sqrt{\frac{M/M_{\odot}}{(R/R_{\odot})^{3}}}\cdot11.66\cyd
\end{equation}
since the radius can be obtained from angular diameter measurements. With the assumed mass of $1.4M_{\odot}$ for $\iota$~Dra the expected frequency splitting is around 0.33\cyd.
But the examination of the frequency splitting needs more effort and much more data for the identification of many modes. Unfortunately, our current data sets are not suitable for this kind of analysis.

\section{Conclusion}

We revised the orbit solution for the companion of $\iota$~Dra. The orbital period is $P=511\,$d, somewhat lower than the value in the discovery paper by \cite{Frink02}. Furthermore there is a linear trend of $-13.8\,$m/s/yr present, possibly caused by another companion.

An excess of power around 3.8\cyd was found independently in two data sets, taken at different times and with very different sampling. The amplitude and the location of the power excess  are consistent with solar-like oscillations.

Our analysis of the short period oscillations indicates a somewhat high stellar mass of $M=2.2M_\odot$, considerably higher than the $1.05M_\odot$ of \cite{Prieto99} or the stellar masses derived  from the Girardi track ($1.2-1.7M_\odot$). However, this is a preliminary result based on limited data.
Assuming $1.4M_\odot$ for the mass $\iota$~Dra yields a minimum mass of $10M_{\mathrm{Jup}}$ for the companion.

Our RV measurements for $\iota$~Dra indicate that it shows multi-periodic stellar oscillations. This means that an asteroseismic analysis can yield an accurate mass. This is best done by measuring the frequency splitting in the
p-mode oscillation spectrum which requires more measurements and better sampling than the data presented here.

\begin{acknowledgements}
We gratefully acknowledge assistance from Debra Fischer and Geoffrey Marcy in obtaining precise radial velocities at Lick Observatory, and we thank the  staff at Lick Observatory for their extraordinary dedication and support. In addition, we would like to thank the CAT observers David Mitchell, Saskia Hekker and Christian Schwab for obtaining some of the observations used in this work.

We thank the anonymous referee for constructive criticism and valuable hints which helped improve this paper.
\end{acknowledgements}

\bibliographystyle{aa}
\bibliography{paper}
\listofobjects
\end{document}